\documentclass[12pt,a4paper]{article}
\usepackage{amsmath,amssymb}
\usepackage{ifpdf}

\ifpdf
  \usepackage[pdftex]{graphicx}
\else
  \usepackage[dvipdfm]{graphicx}
\fi

\newsavebox{\boxa}
\newlength{\bw}

\begin{document}

\thispagestyle{empty}
\begin{flushright}
OU-HET 697
\end{flushright}
\vskip3cm
\begin{center}
{\Large {\bf Statistical model and BPS D4-D2-D0 counting}}
\vskip1.5cm
{\large 
{Takahiro Nishinaka\footnote{nishinaka [at] het.phys.sci.osaka-u.ac.jp}
}
\,and\hspace{2mm} Satoshi Yamaguchi\footnote{yamaguch [at] het.phys.sci.osaka-u.ac.jp}
}
\vskip.5cm
{\it Department of Physics, Graduate School of Science, 
\\
Osaka University, Toyonaka, Osaka 560-0043, Japan}
\end{center}

\vskip2cm
\begin{abstract}
 We construct a statistical model that correctly reproduces the BPS partition function of D4-D2-D0 bound states on the resolved conifold. 
We prove that the known partition function of the BPS indices is reproduced by the counting ``triangular partitions'' problem.
The wall-crossing phenomena in our model are also studied.
\end{abstract}


\newpage

Counting D-brane bound states on a Calabi-Yau three-fold is a longstanding problem in string theory. Since D-branes wrapped on supersymmetric cycles in a Calabi-Yau three-fold can be viewed as BPS particles in the non-compact four dimensions, such a counting problem is relevant for the degeneracy of BPS black holes in $d=4,\,\mathcal{N}=2$ supergravity. In particular, the BPS index of D6-D2-D0 bound states, which is in mathematics called the generalized Donaldson-Thomas invariant, has been studied from various points of view.\footnote{For example, recently, there has been remarkable progress in the study of the wall-crossing phenomena of D6-D2-D0 states and the crystal melting model \cite{Okounkov:2003sp, Iqbal:2003ds, Szendroi:2007nu, Mozgovoy:2008fd, Nagao:2010kx, Nagao, Jafferis:2008uf, Chuang:2008aw, Ooguri:2008yb, Ooguri:2009ri, Dimofte:2009bv, Chuang:2009pd, VanHerck:2009ww, Sulkowski:2009rw, Szabo:2009vw, Krefl:2010sz, Chuang:2010wx, Chuang:2010ii, Aganagic:2010qr}. For the wall-crossing phenomena of D4-D2-D0 states, see \cite{Diaconescu:2007bf, Jafferis:2007ti, Andriyash:2008it, Collinucci:2008ht, Manschot:2009ia, Manschot:2010xp, Nishinaka:2010qk, Nishinaka:2010fh}.} Among them, one of the most interesting results is that the stable BPS D6-D2-D0 bound states on a Calabi-Yau three-fold are in one-to-one correspondence with three-dimensional molten crystals, which was first pointed out on $\mathbb{C}^3$ in \cite{Okounkov:2003sp} and generalized to the resolved conifold case in \cite{Szendroi:2007nu} (See also \cite{Chuang:2008aw}). Its extension to a general toric Calabi-Yau three-fold was given in \cite{Mozgovoy:2008fd,Ooguri:2008yb}. This statistical model description of BPS D6-D2-D0 states provides some insights on the quantum description of geometry in string theory \cite{Iqbal:2003ds, Ooguri:2009ri}.

In this letter, we construct its D4-D2-D0 counterpart, namely, a statistical model that describes BPS bound states of one non-compact D4-brane and arbitrary numbers of D2 and D0-branes on a Calabi-Yau three-fold $X$. We particularly concentrate on the case of the resolved conifold, in which the BPS partition function of D4-D2-D0 states was evaluated in \cite{Nishinaka:2010qk}.\footnote{A generalization to the case of two D4-branes was given in \cite{Nishinaka:2010fh}} The BPS partition function is defined by
\begin{eqnarray}
 \mathcal{Z}(u,v) := \sum_{Q_0,Q_2\in\mathbb{Z}}\Omega(\mathcal{D}+Q_2\beta-Q_0dV)\;u^{Q_0}(-v)^{Q_2},
\nonumber
\end{eqnarray}
where $\Omega(\gamma)$ denotes the BPS index of charge $\gamma$ and $H^{2}(X)\ni \mathcal{D}$ stands for one unit of the non-compact D4-brane charge. The unit D2 and D0-brane charges are denoted by $\beta\in H^{4}(X)$ and $dV\in H^6(X)$ respectively, and therefore $\mathbb{Z}\ni Q_2,\,Q_0$ are the D2 and D0-brane charges of the bound states. Here we set $\beta$ to be dual to the compact two-cycle of the conifold. The Boltzmann weights for D2 and D0-branes are denoted by $-v$ and $u$, respectively.
When putting the non-compact D4-brane on a divisor $C_4=(\text{total space of }\mathcal{O}(-1)\to\mathbb{P}^1)$ and making the compact two-cycle of the conifold shrink to zero size\footnote{In order to keep D2-branes massive, we introduce non-vanishing B-field $B\not\in\mathbb{Z}$ on the compact two-cycle of the conifold. If we change the radius of the compact two-cycle, then the wall-crossing phenomena occur and the BPS partition function might be changed. We will discuss it later.}, the BPS partition function is evaluated as\footnote{See equation (21) in \cite{Nishinaka:2010qk}, where $\mathcal{Z}_{+\infty}(u,v)$ is given in equation (26).}
\begin{eqnarray}
\mathcal{Z}_0(u,v) = \prod_{n=1}^\infty\left(\frac{1}{1-u^n}\right)^{\chi\left(C_4\right)-1}\prod_{m=0}^\infty(1-u^mv),\label{eq:zero-radius}
\end{eqnarray}
where $\chi(C_4)$ represents the Euler characteristic of the divisor $C_4=(\text{total space of }\mathcal{O}(-1)\to\mathbb{P}^1)$. Since the divisor is non-compact, its Euler characteristic cannot be fixed. But here, we simply set $\chi(C_4)=\chi(\mathbb{P}^1)= 2$ by neglecting all the ambiguities due to the non-compactness. Below we construct a two-dimensional {\em triangular partition model} whose partition function is exactly the same as the above D4-D2-D0 partition function \eqref{eq:zero-radius}.

\subsubsection*{Triangular partition model}

\begin{figure}
\begin{center}
\includegraphics[width=6cm]{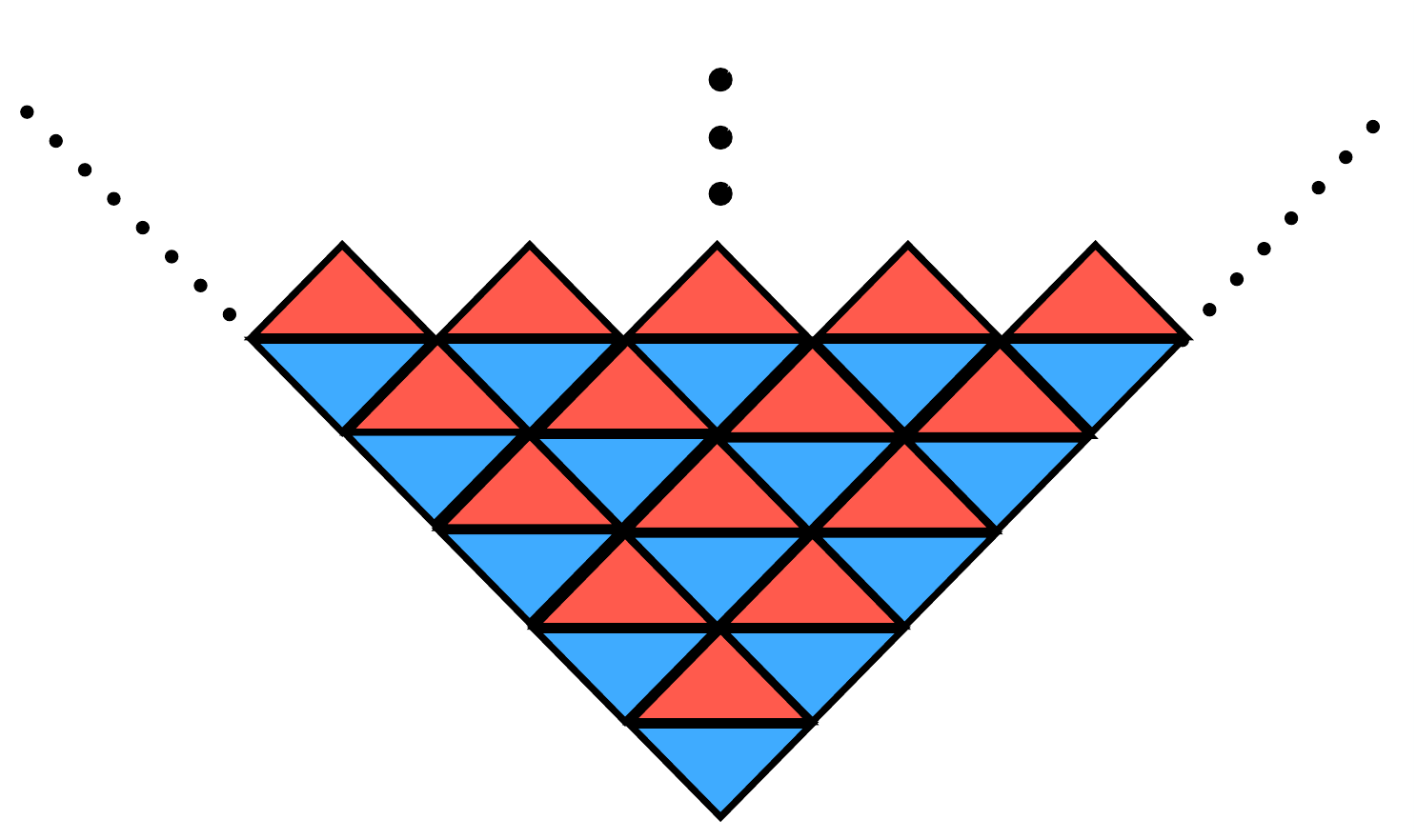}
\qquad\quad
\includegraphics[width=3.5cm]{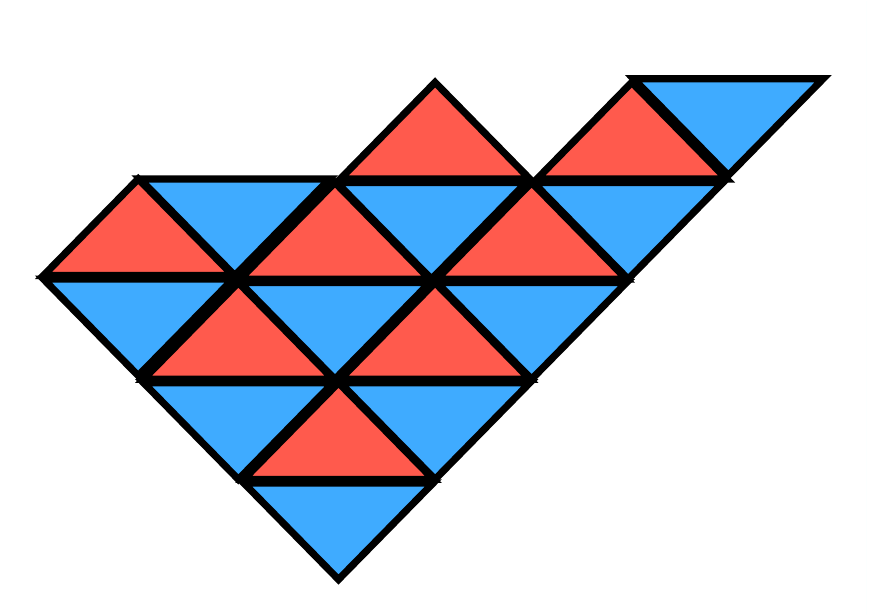}
\caption{Left: The crystal without any excitations. Right: A typical ``triangular partition'' (a set of atoms that can be removed from the crystal).}
\label{fig:crystal1}
\end{center}
\end{figure}

The triangular partition model which we propose is the following.
First, consider a two-dimensional crystal of triangle ``atoms'' as in the left picture of Figure \ref{fig:crystal1}. The crystal is composed of two types of atoms; one is a regular (red) triangle and the other is an inverted (blue) triangle. We associate this crystal to a BPS D4-brane without any D2 and D0-brane charges. From \eqref{eq:zero-radius}, we find that there is only one such state.
Next, we remove some of the atoms from the crystal under the following rules:
\begin{enumerate}
 \item A regular (red) triangle can be removed only if its base is not adjoined to another (blue) triangle.
 \item An inverted (blue) triangle can be removed only if its two sides are not adjoined to other (red) triangles.
\end{enumerate}
Let us call a set of atoms that can be removed from the crystal ``a triangular partition.'' A typical triangular partition is depicted as in the right picture of Figure \ref{fig:crystal1}. Note that the number of the blue atoms in a triangular partition is always greater than or equal to that of the red atoms in the triangular partition. So, if we define
\begin{eqnarray}
 a &=& (\text{the number of red atoms in the triangular partition}),\nonumber
\\
b &=& (\text{the number of blue atoms in the triangular partition}),\nonumber
\end{eqnarray}
then we have $a\leq b$. The partition function of this triangular partition model is defined as
\begin{eqnarray}
 \mathcal{Z}_{\rm triangular} := \sum_{\text{triangular partitions}}x^{a}y^{b},\nonumber
\end{eqnarray}
where $x$ and $y$ express the Boltzmann weights for red and blue removed atoms, respectively.

In the above, we related the pure D4-brane with the vanishing D2 and D0 charges to a crystal without removing atoms. Now we associate each triangular partition to a general D4-D2-D0 bound state, by relating the numbers of removed atoms $a$ and $b$ with the D2 and D0-brane charges. We identify the D0 and D2-brane charges $Q_0, Q_2$ of our D4-D2-D0 states as
\begin{eqnarray}
 Q_0 = a, \qquad Q_2 = -a+b,\label{eq:charges}
\end{eqnarray}
which implies that the chemical potentials $u$ and $v$ for D0 and D2-branes are identified as
\begin{eqnarray}
 u = xy,\qquad v = -y.\label{eq:chemical_potentials}
\end{eqnarray}
The equation \eqref{eq:chemical_potentials} means that we can regard the D0-brane charge as the number of pairs of red and blue atoms.
\begin{figure}
\begin{center}
\includegraphics[width=3.5cm]{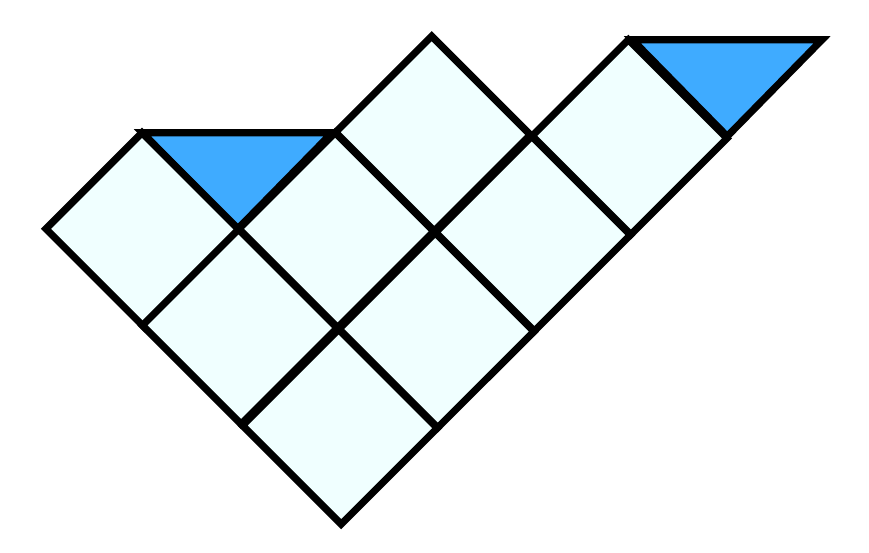}\qquad\quad
\includegraphics[width=4cm]{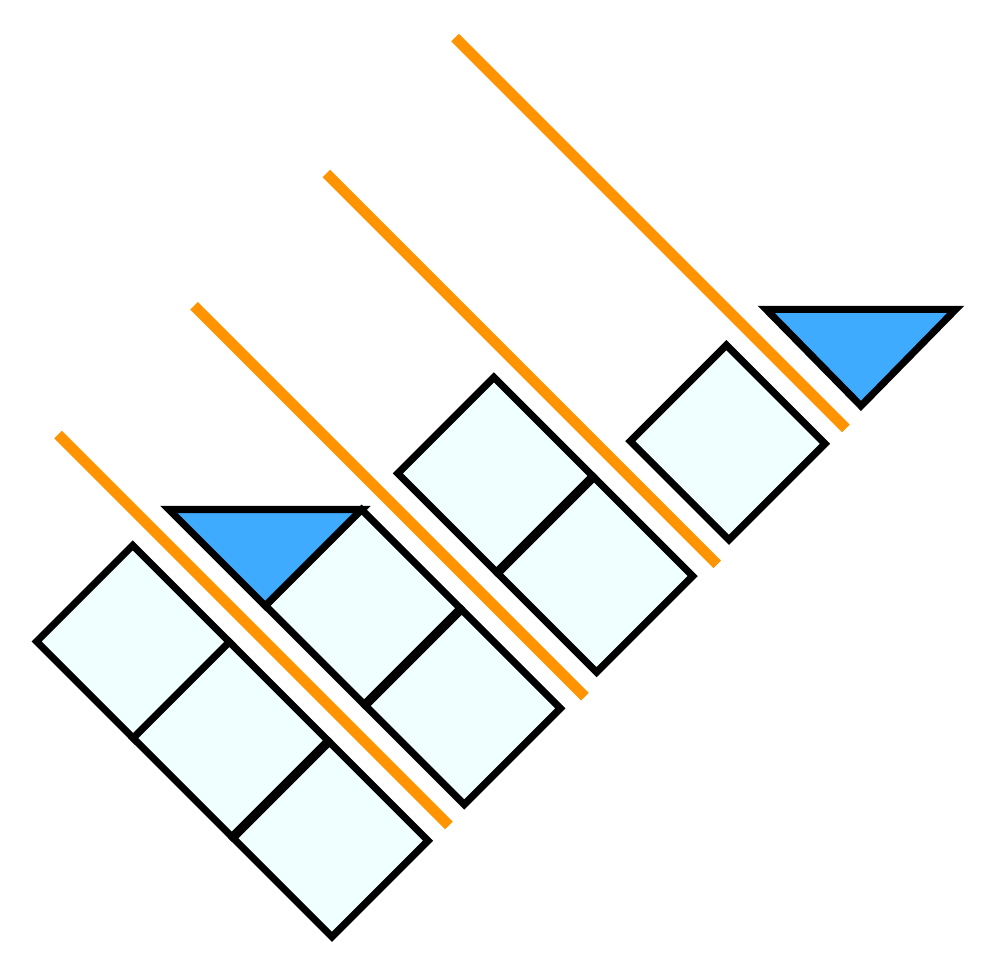}
\caption{Left: Adjacent blue and red atoms can pair up to form a white box, which is regarded as one unit of the D0-brane charge. The number of the remaining blue atoms is identified with the D2-brane charge. The above example has the D0-charge $8$ and the D2-charge $2$. Right: A set of removed atoms (a triangular partition) can be divided into towers of white boxes. Some towers have one blue triangle at their edges while the others have no blue triangles and only contain white boxes.}
\label{fig:divide_atoms}
\end{center}
\end{figure}
Adjacent blue and red atoms can pair up to form a white box, and the number of such boxes is identified with the D0-brane charge (See the left picture of Figure \ref{fig:divide_atoms}).\footnote{Of course, we here make as many white boxes as possible, so that all the red atoms are paired with blue atoms.} Since there are more blue atoms than red ones, \eqref{eq:chemical_potentials} implies that the number of the remaining blue atoms should be identified with the D2-brane charge. Note that we can always make the white boxes so that all the remaining blue atoms are at the upper edge of the triangular partition.

\subsubsection*{Reproducing the D4-D2-D0 partition function}

\begin{figure}
\begin{center}
\includegraphics[width=5cm]{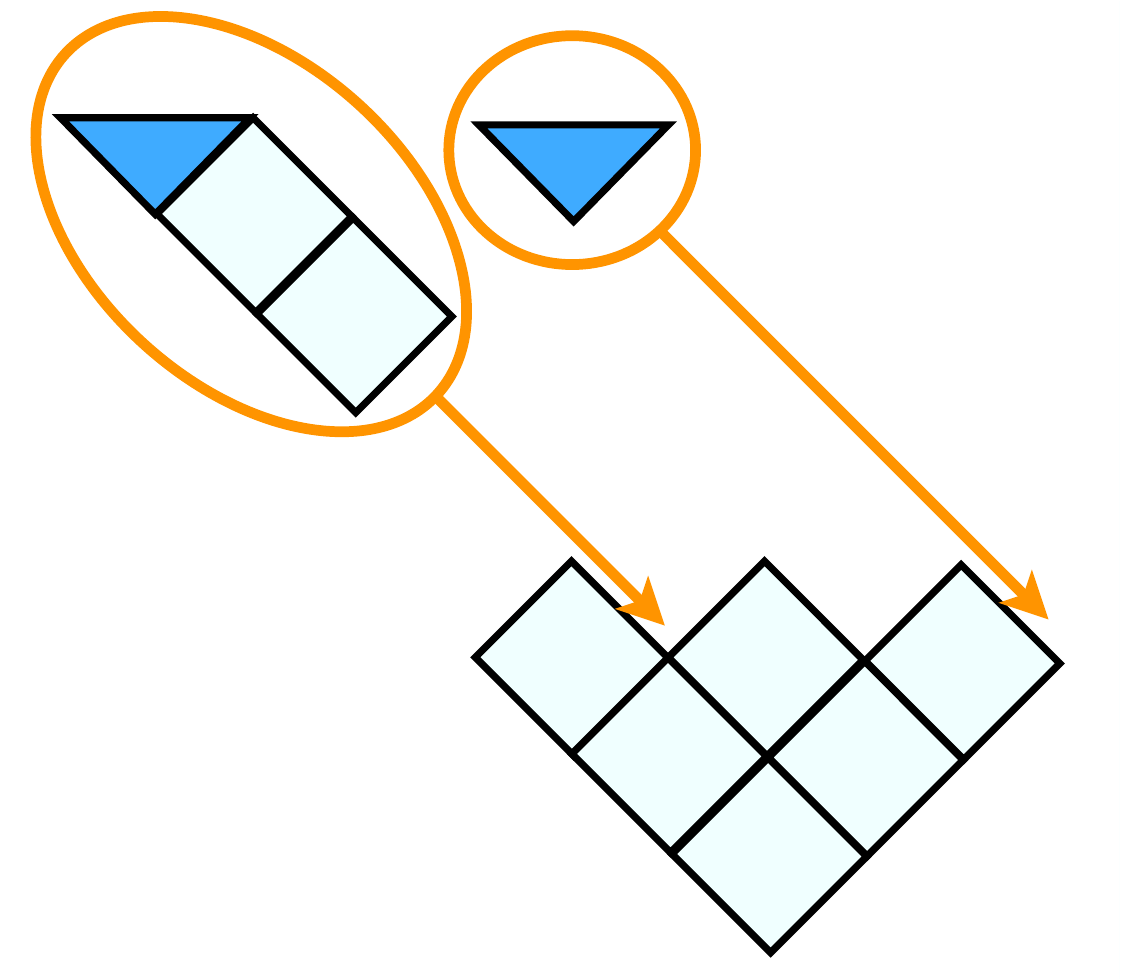}
\caption{The triangular partitions are in one-to-one correspondence with the ``fermionic'' insertions of the form \eqref{eq:insertion} into usual two-dimensional Young diagrams.}
\label{fig:insertion}
\end{center}
\end{figure}

We now show that with the identification \eqref{eq:charges} the triangular partition model correctly reproduces the D4-D2-D0 partition function \eqref{eq:zero-radius}, that is,
\begin{eqnarray}
\mathcal{Z}_{\rm triangular}
\;:= \sum_{\text{triangular partitions}}x^{a}y^{b}
\;=\; \prod_{n=1}^\infty\left(\frac{1}{1-u^n}\right)\prod_{m=0}^\infty\left(1-u^mv\right).\label{eq:reconstruction}
\end{eqnarray}
To see this, we divide a triangular partition as in the right picture of Figure \ref{fig:divide_atoms}.
We cut up the triangular partition into towers of white boxes.
Some of the towers have one blue triangle at their edges while the others only contain white boxes. By collecting the latter, we can construct a usual Young diagram. On the other hand, the former is
labeled by the number of white boxes $m\geq 0$. An example for $m=3$ is
\sbox{\boxa}{\includegraphics[width=2.2cm]{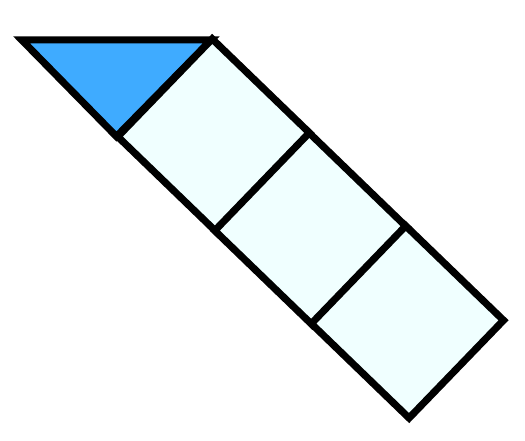}}
\settowidth{\bw}{\usebox{\boxa}}
\begin{eqnarray}
\parbox{\bw}{\usebox{\boxa}}\label{eq:insertion}.
\end{eqnarray}
In other words, a triangular partition is reconstructed by inserting some towers of the form of \eqref{eq:insertion} into a Young diagram as in Figure \ref{fig:insertion}.
Because of the rules of removing atoms, the insertion of the tower for a fixed $m$ cannot be made more than once. So these insertions are ``fermionic.'' Furthermore, for any $m\geq 0$, the position of the insertion is determined unambiguously so that the resulting atoms can be removed from the crystal under our rules.\footnote{To be more precise, the insertion of a tower with $m$ white boxes and one blue triangle should be made so that all towers with $m_1$ white boxes for $m_1\leq m$ exist in the right side of the insertion while towers with $m_2$ white boxes for $m_2>m$ exist in its left side.} This implies that {\em the triangular partitions are in one-to-one correspondence with the ``fermionic'' insertions of the towers like \eqref{eq:insertion} into two-dimensional Young diagrams.}

According to the identification \eqref{eq:chemical_potentials}, an insertion of the tower with one blue triangle and $m$ white boxes has the D0-charge $m$ and D2-charge one, and therefore its contribution to the partition function is $-u^mv$. 
All such insertions of the towers are taken into account by multiplying
\begin{eqnarray}
 \prod_{m=0}^\infty\left(1-u^mv\right).\nonumber
\end{eqnarray}
On the other hand, according to the identification rules \eqref{eq:chemical_potentials}, summing up all the two-dimensional Young diagrams gives
\begin{eqnarray}
 \prod_{n=1}^\infty\left(\frac{1}{1-u^n}\right).\nonumber
\end{eqnarray}
Thus, {\em collecting all the insertions into all the Young diagrams reproduces the right-hand side of \eqref{eq:reconstruction}.}

The above argument implies that our triangular partition model correctly reproduces \eqref{eq:zero-radius} which is the D4-D2-D0 partition function in the zero-radius limit of the compact two-cycle of the conifold.
Namely, the stable BPS D4-D2-D0 bound states on the conifold singularity are in one-to-one correspondence with the triangular partitions.

\subsubsection*{Discussions}

In this letter, we have constructed a statistical model that can reproduce the BPS partition function of the D4-D2-D0 states on the conifold singularity. We now briefly discuss the wall-crossing phenomena in our model.
Let $z$ be the complexified K\"ahler parameter for the compact two-cycle of the conifold, where ${\rm Re}\,z$ and $\left|{\rm Im}\,z\right|$ represent the B-field and the area of the compact two-cycle, respectively. When we change the K\"ahler parameter $z$, the BPS index might be changed at some codimension one subspace in the moduli space. This is called the wall-crossing phenomenon. The wall-crossing of our D4-D2-D0 system was studied in \cite{Nishinaka:2010qk}, and it was shown that the BPS partition function is given by
\begin{eqnarray}
 \mathcal{Z}_k(u,v) \;=\; \prod_{n=1}^\infty\left(\frac{1}{1-u^n}\right)\prod_{m=k}^\infty\left(1-u^mv\right),\nonumber
\end{eqnarray}
when we set $z=1/2+ia$ for $a\leq 0$.\footnote{We define $z$ so that our D4-brane is wrapped on $\mathcal{O}(-1)\to\mathbb{P}^1$ in the moduli region of ${\rm Im}\,z>0$. When ${\rm Im}\,z=0$, the flop transition of the conifold occurs and the topology of the four-cycle wrapped by the D4-brane is changed. Then, in the regime ${\rm Im}\,z<0$, our D4-brane is localized on $\mathbb{P}^1$ and extended along the whole fiber direction of the conifold $\mathcal{O}(-1)\oplus\mathcal{O}(-1)\to\mathbb{P}^1$.} Here the integer $k\geq 0$ represents the number of wall-crossings that occur when the moduli $z$ is moved from ${\rm Im}\,z=0$ to ${\rm Im}\,z = a$. We can easily construct this wall-crossed version of the partition function from \eqref{eq:zero-radius} by changing the variables as
\begin{eqnarray}
 u\to u,\quad v \to u^kv.\label{eq:change-variable}
\end{eqnarray}
Thus, essentially, the same triangular partition model description is applicable to all the chambers including $z=1/2 + ia$ for $a\leq 0$, by taking into account the change of the variables \eqref{eq:change-variable}. The generalization to the moduli region for $a>0$ is left to future work. 

Another interesting future problem will be the study on the thermodynamic limit of our model. In \cite{Okounkov:2003sp,Iqbal:2003ds,Ooguri:2009ri}, it was shown that the thermodynamic limit of the three-dimensional crystal melting model for D6-D2-D0 system gives rise to a smooth mirror Calabi-Yau geometry, which means that the crystal melting model could be a quantum description of geometry in string theory. Correspondingly, the thermodynamic limit of our triangular partition model is expected to provide some information on the mirror of the divisor wrapped by the D4-brane.

\section*{Acknowledgments}

We would like to thank Takahiro Kubota for many illuminating discussions, important comments and suggestions.
We also wish to thank Yu Nakayama, Hirosi Ooguri and Piotr Su\l kowski for many useful discussions.
T.N. was supported in part by JSPS Research Fellowship for Young Scientists. 
S.Y. was supported in part by KAKENHI 22740165.

\bibliography{ref}

\providecommand{\href}[2]{#2}\begingroup\raggedright\begin{thebibliography}{10}

\bibitem{Okounkov:2003sp}
A.~Okounkov, N.~Reshetikhin, and C.~Vafa, ``{Quantum Calabi-Yau and classical
  crystals},''
\href{http://arxiv.org/abs/hep-th/0309208}{{\ttfamily arXiv:hep-th/0309208}}.

\bibitem{Iqbal:2003ds}
A.~Iqbal, N.~Nekrasov, A.~Okounkov, and C.~Vafa, ``{Quantum foam and
  topological strings},''
  \href{http://dx.doi.org/10.1088/1126-6708/2008/04/011}{{\em JHEP} {\bfseries
  04} (2008) 011},
\href{http://arxiv.org/abs/hep-th/0312022}{{\ttfamily arXiv:hep-th/0312022}}.

\bibitem{Szendroi:2007nu}
B.~Szendroi, ``{Non-commutative Donaldson-Thomas theory and the conifold},''
  \href{http://dx.doi.org/10.2140/gt.2008.12.1171}{{\em Geom. Topol.}
  {\bfseries 12} (2008) 1171--1202},
\href{http://arxiv.org/abs/0705.3419}{{\ttfamily arXiv:0705.3419 [math.AG]}}.

\bibitem{Mozgovoy:2008fd}
S.~Mozgovoy and M.~Reineke, ``{On the noncommutative Donaldson-Thomas
  invariants arising from brane tilings},''
\href{http://arxiv.org/abs/0809.0117}{{\ttfamily arXiv:0809.0117 [math.AG]}}.

\bibitem{Nagao:2010kx}
K.~Nagao and H.~Nakajima, ``{Counting invariant of perverse coherent sheaves
  and its wall-crossing},''
\href{http://arxiv.org/abs/0809.2992}{{\ttfamily arXiv:0809.2992 [math.AG]}}.

\bibitem{Nagao}
K.~Nagao, ``{Derived categories of small toric Calabi-Yau 3-folds and counting
  invariants},'' \href{http://arxiv.org/abs/0809.2994}{{\ttfamily
  arXiv:0809.2994 [math.AG]}}.

\bibitem{Jafferis:2008uf}
D.~L. Jafferis and G.~W. Moore, ``{Wall crossing in local Calabi Yau
  manifolds},''
\href{http://arxiv.org/abs/0810.4909}{{\ttfamily arXiv:0810.4909 [hep-th]}}.

\bibitem{Chuang:2008aw}
W.-y. Chuang and D.~L. Jafferis, ``{Wall Crossing of BPS States on the Conifold
  from Seiberg Duality and Pyramid Partitions},''
  \href{http://dx.doi.org/10.1007/s00220-009-0832-2}{{\em Commun. Math. Phys.}
  {\bfseries 292} (2009) 285--301},
\href{http://arxiv.org/abs/0810.5072}{{\ttfamily arXiv:0810.5072 [hep-th]}}.

\bibitem{Ooguri:2008yb}
H.~Ooguri and M.~Yamazaki, ``{Crystal Melting and Toric Calabi-Yau
  Manifolds},'' \href{http://dx.doi.org/10.1007/s00220-009-0836-y}{{\em Commun.
  Math. Phys.} {\bfseries 292} (2009) 179--199},
\href{http://arxiv.org/abs/0811.2801}{{\ttfamily arXiv:0811.2801 [hep-th]}}.

\bibitem{Ooguri:2009ri}
H.~Ooguri and M.~Yamazaki, ``{Emergent Calabi-Yau Geometry},''
  \href{http://dx.doi.org/10.1103/PhysRevLett.102.161601}{{\em Phys. Rev.
  Lett.} {\bfseries 102} (2009) 161601},
\href{http://arxiv.org/abs/0902.3996}{{\ttfamily arXiv:0902.3996 [hep-th]}}.

\bibitem{Dimofte:2009bv}
T.~Dimofte and S.~Gukov, ``{Refined, Motivic, and Quantum},''
  \href{http://dx.doi.org/10.1007/s11005-009-0357-9}{{\em Lett. Math. Phys.}
  {\bfseries 91} (2010) 1},
\href{http://arxiv.org/abs/0904.1420}{{\ttfamily arXiv:0904.1420 [hep-th]}}.

\bibitem{Chuang:2009pd}
W.-y. Chuang and G.~Pan, ``{BPS State Counting in Local Obstructed Curves from
  Quiver Theory and Seiberg Duality},'' {\em J. Math. Phys.} {\bfseries 51}
  (2010) 052305,
\href{http://arxiv.org/abs/0908.0360}{{\ttfamily arXiv:0908.0360 [hep-th]}}.

\bibitem{VanHerck:2009ww}
W.~Van~Herck and T.~Wyder, ``{Black Hole Meiosis},''
  \href{http://dx.doi.org/10.1007/JHEP04(2010)047}{{\em JHEP} {\bfseries 04}
  (2010) 047},
\href{http://arxiv.org/abs/0909.0508}{{\ttfamily arXiv:0909.0508 [hep-th]}}.

\bibitem{Sulkowski:2009rw}
P.~Sulkowski, ``{Wall-crossing, free fermions and crystal melting},''
\href{http://arxiv.org/abs/0910.5485}{{\ttfamily arXiv:0910.5485 [hep-th]}}.

\bibitem{Szabo:2009vw}
R.~J. Szabo, ``{Instantons, Topological Strings and Enumerative Geometry},''
  \href{http://dx.doi.org/10.1155/2010/107857}{{\em Adv. Math. Phys.}
  {\bfseries 2010} (2010) 107857},
\href{http://arxiv.org/abs/0912.1509}{{\ttfamily arXiv:0912.1509 [hep-th]}}.

\bibitem{Krefl:2010sz}
D.~Krefl, ``{Wall Crossing Phenomenology of Orientifolds},''
\href{http://arxiv.org/abs/1001.5031}{{\ttfamily arXiv:1001.5031 [hep-th]}}.

\bibitem{Chuang:2010wx}
W.-y. Chuang, D.-E. Diaconescu, and G.~Pan, ``{Rank Two ADHM Invariants and
  Wallcrossing},''
\href{http://arxiv.org/abs/1002.0579}{{\ttfamily arXiv:1002.0579 [math.AG]}}.

\bibitem{Chuang:2010ii}
W.-y. Chuang, D.-E. Diaconescu, and G.~Pan, ``{Wallcrossing and Cohomology of
  The Moduli Space of Hitchin Pairs},''
\href{http://arxiv.org/abs/1004.4195}{{\ttfamily arXiv:1004.4195 [math.AG]}}.

\bibitem{Aganagic:2010qr}
M.~Aganagic and K.~Schaeffer, ``{Wall Crossing, Quivers and Crystals},''
\href{http://arxiv.org/abs/1006.2113}{{\ttfamily arXiv:1006.2113 [hep-th]}}.

\bibitem{Diaconescu:2007bf}
E.~Diaconescu and G.~W. Moore, ``{Crossing the Wall: Branes vs. Bundles},''
\href{http://arxiv.org/abs/0706.3193}{{\ttfamily arXiv:0706.3193 [hep-th]}}.

\bibitem{Jafferis:2007ti}
D.~L. Jafferis and N.~Saulina, ``{Fragmenting D4 branes and coupled q-deformed
  Yang Mills},''
\href{http://arxiv.org/abs/0710.0648}{{\ttfamily arXiv:0710.0648 [hep-th]}}.

\bibitem{Andriyash:2008it}
E.~Andriyash and G.~W. Moore, ``{Ample D4-D2-D0 Decay},''
\href{http://arxiv.org/abs/0806.4960}{{\ttfamily arXiv:0806.4960 [hep-th]}}.

\bibitem{Collinucci:2008ht}
A.~Collinucci and T.~Wyder, ``{The elliptic genus from split flows and
  Donaldson-Thomas invariants},''
  \href{http://dx.doi.org/10.1007/JHEP05(2010)081}{{\em JHEP} {\bfseries 05}
  (2010) 081},
\href{http://arxiv.org/abs/0810.4301}{{\ttfamily arXiv:0810.4301 [hep-th]}}.

\bibitem{Manschot:2009ia}
J.~Manschot, ``{Stability and duality in N=2 supergravity},''
  \href{http://dx.doi.org/10.1007/s00220-010-1104-x}{{\em Commun. Math. Phys.}
  {\bfseries 299} (2010) 651--676},
\href{http://arxiv.org/abs/0906.1767}{{\ttfamily arXiv:0906.1767 [hep-th]}}.

\bibitem{Manschot:2010xp}
J.~Manschot, ``{Wall-crossing of D4-branes using flow trees},''
\href{http://arxiv.org/abs/1003.1570}{{\ttfamily arXiv:1003.1570 [hep-th]}}.

\bibitem{Nishinaka:2010qk}
T.~Nishinaka and S.~Yamaguchi, ``{Wall-crossing of D4-D2-D0 and flop of the
  conifold},'' \href{http://dx.doi.org/10.1007/JHEP09(2010)026}{{\em JHEP}
  {\bfseries 09} (2010) 026}, \href{http://arxiv.org/abs/1007.2731}{{\ttfamily
  arXiv:1007.2731 [hep-th]}}.

\bibitem{Nishinaka:2010fh}
T.~Nishinaka, ``{Multiple D4-D2-D0 on the Conifold and Wall-crossing with the
  Flop},'' \href{http://arxiv.org/abs/1010.6002}{{\ttfamily arXiv:1010.6002
  [hep-th]}}.

\end{thebibliography}\endgroup

\end{document}